# Self-Assembly of Core-Shell Hybrid Nanoparticles by Directional Crystallization of Grafted Polymers

A. Nabiyan,[a,b,c] A. Muttathukattil,[d] F. Tomazic,[d] D. Pretzel,[a,b] Ulrich. S. Schubert,[a,b] M. Engel,[d]* F. H. Schacher [a,b,c]*

[a]Jena Center for Soft Matter (JCSM), Friedrich Schiller University Jena, Philosophenweg 7, D-07743 Jena, Germany.

[b]Institute of Organic Chemistry and Macromolecular Chemistry (IOMC), Friedrich-Schiller University Jena, Lessingstraße 8, D-07743 Jena, Germany.

[c]Center for Energy and Environmental Chemistry (CEEC), Friedrich-Schiller University Jena, Philosophenweg 7, D-07743 Jena, Germany.

[d]Institute for Multiscale Simulation, IZNF, Friedrich-Alexander-Universität Erlangen-Nürnberg, Cauerstrasse 3, 91058 Erlangen, Germany.

Keywords: self-assembly, crystallization, hybrid materials, isotropic and anisotropic nanomaterials.

**ABSTRACT:** Nanoparticle self-assembly is an efficient bottom-up strategy for the creation of nanostructures. In the standard approach, ligands are grafted on the surfaces of nanoparticles to keep them separated and control interparticle interactions. Ligands then remain secondary and usually are not expected to order significantly during superstructure formation. Here, we investigate how ligands can play a more primary role in the formation of inorganic-organic hybrid materials. We graft poly(2-*iso*-propyl-2-oxazoline) (P*i*PrOx) as a crystallizable shell onto $SiO_2$ nanoparticles. By varying the P*i*PrOx grafting density, solution stability, and nanoparticle aggregation behavior can be controlled. Upon prolonged heating, anisotropic nanostructures form in conjunction with the crystallization of the ligands. Self-assembly of hybrid P*i*PrOx@$SiO_2$ (shell@core) nanoparticles proceeds in two steps: First, rapid formation of amorphous aggregates *via* gelation, mediated by the interaction between nanoparticles through grafted polymers; second, slow radial growth of fibers *via* directional crystallization, governed by the incorporation of crystalline ribbons formed from unbound polymers coupling to the grafted polymer shell. Our work reveals how crystallization-driven self-assembly of ligands can create intricate hybrid nanostructures.

## Introduction

Ensembles of organic and inorganic nanoparticles (NPs) have collective properties that are relevant for applications in functional devices.[1,2] For instance, coupling plasmon resonances of NPs enhances fluorescence and Raman scattering with application in photothermal therapy and photocatalysis.[3–6] Various self-assembly strategies have been reported: Self-assembly at interfaces,[7,8] by external fields,[9,10] using soft or hard templating,[11–13] in solution using molecular ligands, *i.e.*, small molecules, polymers, and DNA.[1,6,14–16] Among those strategies, decorating the surface of NPs with hydrophilic/hydrophobic polymers drew particular attention because of the manifold possibilities in varying ligand chemistry.[17–23]

Self-assembly of polymer-grafted (or 'hairy') NPs[20] can utilize anisotropic building blocks[24,25] or profit from the spontaneous emergence of anisotropic structures with spherical building blocks.[13,19,23,26] The latter class is intriguing from the perspectives of cost, scale-up, and processing. For instance, silica nanospheres functionalized by nonionic poly(ethylene oxide)-*block*-poly(propylene oxide)-*block*-poly(ethylene oxide) ligands assemble into short chains by pH adjustment,[27] and the monomer sequence in grafted copolymers is a tuning parameter to control the size and shape of assembled nanoclusters.[28] Furthermore, poly(L-glutamic acid)-grafted gold NPs associate directionally for the cooperative growth of supramolecular chains,[29] and the dimension of gold NP clusters coated by polystyrene is controlled by temperature and incubation time.[6] Yet, despite significant progress in the assembly of polymer-grafted NPs,[2] the quantitative prediction of superstructures remains a challenge, and representative model systems and better resolution of the formation process kinetics are desirable.

Most studies of polymer-grafted NP assembly have been focusing on weakly-ordered ligand shells dominated by van der Waals or electrostatic interactions. A different approach is ligands with semicrystalline polymer segments that, depending on temperature and solvent quality, undergo directional crystallization.[30,31] In this case, the term crystallization-driven self-assembly (CDSA) was established. One known hydrophilic polymer that undergoes CDSA is poly(2-*iso*-propyl-2-oxazoline) (P*i*PrOx).[32–34] P*i*PrOx is a thermoresponsive hydrophilic polymer that crystallizes upon prolonged heating above the cloud point temperature.[35,36] While the CDSA of P*i*PrOx as a part of block copolymers with additional hydrophilic or hydrophobic blocks is well established,[37–39] CDSA of inorganic NPs with P*i*PrOx ligands has not been explored.

This work investigates spherical $SiO_2$ NPs uniformly grafted with P*i*PrOx. We demonstrate that hybrid P*i*PrOx@$SiO_2$ (organic-inorganic shell@core) NPs robustly self-assemble into a variety of anisotropic superstructures during prolonged heating while the shell undergoes CDSA. Our experiments suggest that the self-assembly process reflects a balance between the solution stability of the core-shell NPs and the tunable directional crystallization of the shell. Our molecular dynamics simulations resolve details of the self-assembly mechanism and explain how the directional crystallization of PiPrOx guides the formation of the anisotropic structures observed in the experiments.

## Methods

CDSA has been established as a method for the formation of anisotropic nanostructures from crystalline-coil block copolymers. Here, we investigate whether spherical building blocks featuring a polymeric shell capable of undergoing CDSA can also be employed in similar processes by grafting semi-crystalline PiPrOx polymer ligands to the surface of different $SiO_2$ NPs.

### Synthesis of PiPrOx-Si(OEt)$_3$

PiPrOx was covalently bonded to $SiO_2$ NPs following the *grafting to-approach*" (Figure 1A).[40] For this approach, two batches of PiPrOx with 30 (PiPrOx$_{30}$) and 100 (PiPrOx$_{100}$) repeating units were prepared by cationic ring-opening polymerization of 2-*iso*-propyl-2-oxazoline. Polymerization was initiated by methyl *p*-toluene sulfonate (MeTos, Figure 1A). The resulting PiPrOx shows narrow dispersity (Đ) and good agreement between theoretical and empirical molar mass (Mn) derived from size exclusion chromatography (SEC) and $^1$H nuclear magnetic resonance (NMR) spectroscopy (Table 1). PiPrOx with Mn ranging from 5000 to 11500 g·mol$^{-1}$ and with Đ from 1.03 to 1.1 was obtained in a well-controlled manner. Afterward, to generate a hydroxyl end group, the reaction was terminated with an aqueous $NaHCO_3$ solution. The resulting PiPrOx-OH were subsequently covalently grafted to $SiO_2$ NPs after end-group functionalization using IPTES (Figure 1A). The obtained PiPrOx-Si(OEt)$_3$ were evaluated for synthesis and functionalization success again by SEC and $^1$H NMR spectroscopy. $^1$H NMR spectra of PiPrOx-Si(OEt)$_3$ are illustrated in Figure 1C. The CH$_2$ signals of the linker (9, 7, and 8) and the carbamate proton (6) are both present. In addition, the SEC trace in Figure 1B shows a slight shift to lower elution volume, confirming successful end-group modification.

### Table 1. Characterization of the PiPrOx Polymers.

| Polymer | M:I[a] | DP[b] | Mn (g/mol)[c] | Đ[c] |
|---|---|---|---|---|
| PiPrOx$_{30}$ | 30 | 30 | 4000 | 1.03 |
| PiPrOx$_{100}$ | 100 | 100 | 11300 | 1.10 |

[a] Monomer-to-initiator ratio used in polymerizations, [b] Degree of polymerization (DP) calculated by comparison of peak integrals from the initiating methyl group at the α chain and the polymer backbone. [c] Molar mass (Mn) and dispersity (Đ) obtained from SEC (DMAC with PMMA calibration).

### Grafting of PiPrOx-Si(OEt)$_3$ onto $SiO_2$ Nanoparticles

$SiO_2$ NPs as cores were prepared with two different sizes in ethanol using the Stöber method.[41] Hydrodynamic diameters (R$_h$) were determined using a combination of transmission electron microscopy (TEM) and dynamic light scattering (DLS) as 15 nm and 25 nm (Figure 1D and Figure S1).

For the attachment of a PiPrOx shell, a solution of PiPrOx-Si(OEt)$_3$ in ethanol was slowly added to the obtained ammoniacalic $SiO_2$ NPs under an Ar atmosphere. After stirring for 24 hours at room temperature, PiPrOx@SiO$_2$ particles were dialyzed against water for three days to remove free PiPrOx-Si(OEt)$_3$. The obtained core-shell hybrid NPs were evaluated again with DLS (Figure S1C, D). R$_h$ increased from 15 to 23 nm, and from 25 to 36 nm, indicating the formation of a PiPrOx-Si(OEt)$_3$ shell. Morphology slightly changed if compared to the pristine NP, which is demonstrated by the fact that the particle surface appears less smooth and a thin gray layer can be observed (Figure 1E).

Further evidence from attenuated total reflectance (ATR) infrared (IR) spectroscopy confirmed the success of surface functionalization. In ATR spectra (Figure S2), besides the three absorption bands at 1100 cm$^{-1}$ (Si–O stretching), 953 cm$^{-1}$ (Si–OH bending), and 800 cm$^{-1}$ (Si–O–Si bending) from bare $SiO_2$, the characteristic absorption of the carbonyl groups from the attached PiPrOx@SiO$_2$ on the NP surface was observed at 1720 cm$^{-1}$.

To quantify the amount of grafted PiPrOx, thermogravimetric analysis (TGA) of pristine $SiO_2$ and PiPrOx@SiO$_2$ was performed. To exclude the influence of merely adsorbed polymer chains, TGA values after 3, 5, and 7 days against dialysis were compared and found to be comparable. As shown in Figure S3, an increased mass loss was observed for PiPrOx@SiO$_2$ compared to pristine $SiO_2$, confirming successful grafting. The grafting density (σ) can be calculated using the core size of the $SiO_2$ NP and the molar masses of the grafted polymers to determine the amount of PiPrOx to $SiO_2$ (Equation 1 in Supporting Information).[40] Table 2 shows σ and the calculated number of PiPrOx chains at the $SiO_2$ surface for a variety of PiPrOx@SiO$_2$ core-shell hybrid NPs. To later investigate the solution properties of PiPrOx@SiO$_2$ and to evaluate the role of the PiPrOx or $SiO_2$ in the solution behavior, σ was varied between 0.11 and 0.47 chains/nm$^2$.

### Table 2. Characterization of Synthesized PiPrOx@SiO$_2$ Hybrid NPs.

| Sample | [b] (wt%) | [c] No. | [d] σ |
|---|---|---|---|
| PiPrOx$_{100}$@SiO$_2$ [15 nm][a] | 17 | 1800 | 0.11 |
| PiPrOx$_{100}$@SiO$_2$ [15 nm][a] | 23 | 2600 | 0.18 |
| PiPrOx$_{100}$@SiO$_2$ [15 nm][a] | 37.8 | 5100 | 0.35 |
| PiPrOx$_{100}$@SiO$_2$ [15 nm][a] | 44.4 | 6900 | 0.47 |
| PiPrOx$_{100}$@SiO$_2$ [25 nm][a] | 21.5 | 1900 | 0.24 |
| PiPrOx$_{30}$@SiO$_2$ [25 nm][a] | 15.3 | 5400 | 0.70 |

[a] Average hydrodynamic radii of the core determined by DLS. [b] Weight loss of core-shell hybrids determined by TGA. [c] Average number of PiPrOx chains per $SiO_2$ NP (chains/$SiO_2$). [d] Grafting density (chains/nm$^2$).

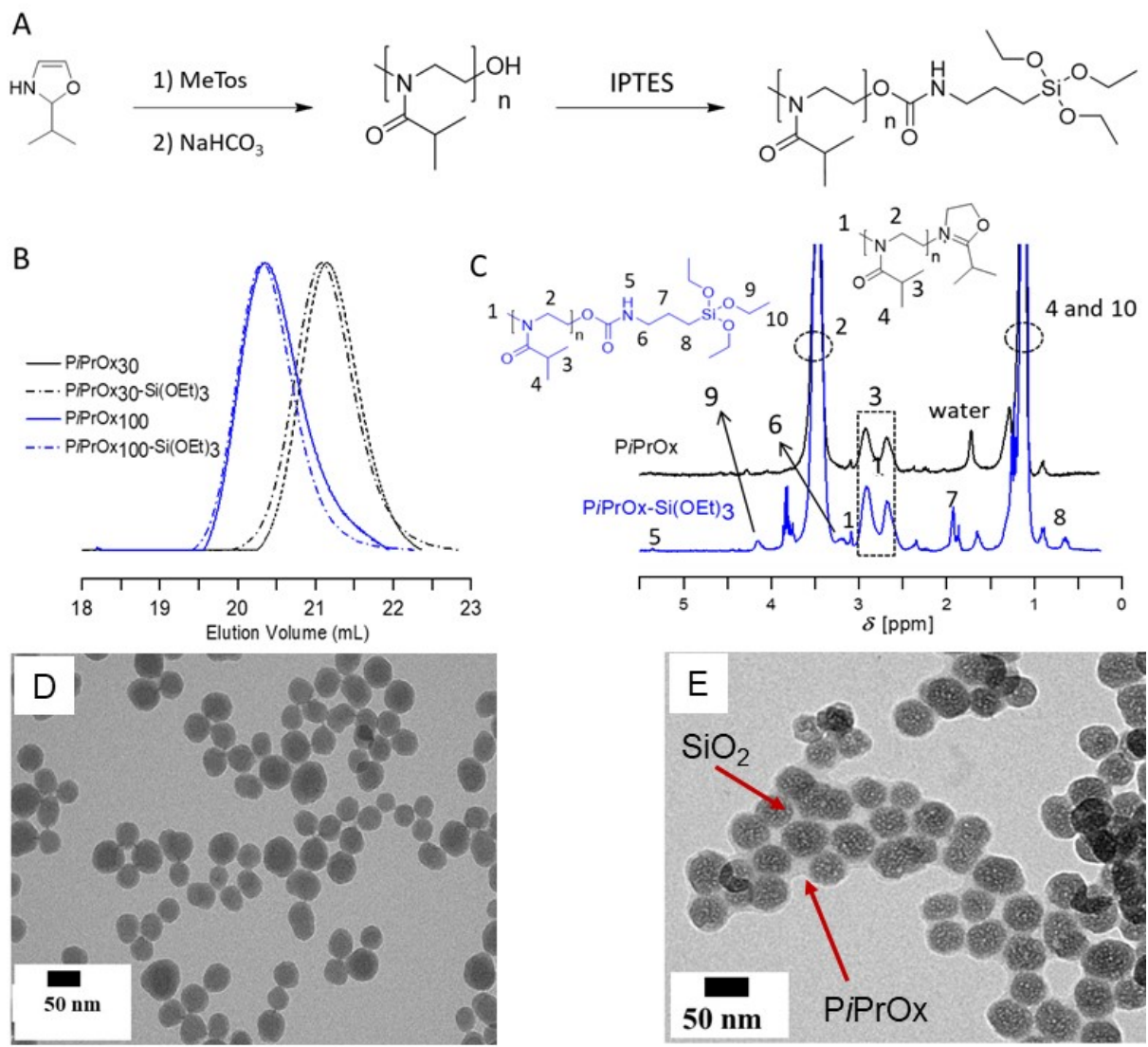

**Figure 1.** (A) Schematic of P*i*PrOx synthesis and subsequent end group modification with IPTES. (B) SEC elution traces (DMAC, LiCl) of P*i*PrOx and P*i*PrOx-Si(OEt)$_3$ with a DP of 30 and 100. (C) $^1$H NMR (CDCl$_3$, 300 MHz) of P*i*PrOx$_{100}$ and P*i*PrOx$_{100}$-Si(OEt)$_3$. TEM micrographs of (D) pristine SiO$_2$ ($R_h$ = 25 nm) and (E) P*i*PrOx$_{100}$@SiO$_2$ ($R_h$ = 36 nm)

.

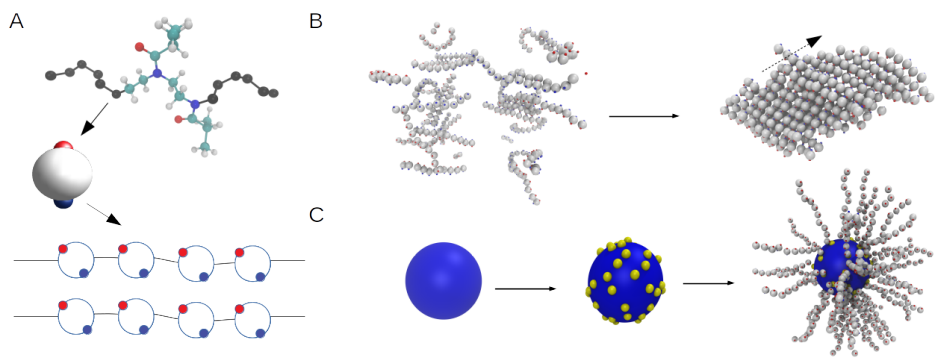

**Figure 2.** (A): Schematic of coarse-grained (CG) model for P*i*PrOx as a chain of patchy particles. Large white beads are parent beads representing two P*i*PrOx monomers each. Small blue and red beads on the surface of the parent bead represent directional dipolar interaction patches due to amide groups facing in diametrically opposed directions. (B): P*i*PrOx chains tend to self-assemble directionally into ribbon-like structures. (C): SiO$_2$ NPs are represented by spheres (blue). Binding sites for P*i*PrOx chains (yellow) are distributed randomly over the surface.

## Molecular Dynamics Simulation Methods

Coarse-grained (CG) modeling of P*i*PrOx@SiO$_2$ NPs was employed to reach a large system size and to capture the self-assembly process. P*i*PrOx polymers were mapped to chains of patchy beads that are then attached to SiO$_2$ surfaces (Figure 2). At the molecular scale, two types of interactions dominate the crystallization of P*i*PrOx.[35] These interactions are the directional dipolar interactions between amide groups and the non-directional hydrophobic interaction between the isopropyl side-chain. In our CG model, one bead represents two monomers of the P*i*PrOx polymer, which is the repeating unit along the backbone in the crystalline state (Figure 2A). The directionality of the amide dipolar interaction was represented by two patches on each CG bead. Such an addition is inspired by earlier simulations where CG beads mimic directional interactions such as hydrogen bonds in cellulose and DNA-based systems.[42,43] To incorporate the desired structural features, such as the geometrical arrangement of monomers within a single chain, and model parameters, such as bending stiffness, we performed all-atom simulations (AA) of P*i*PrOx polymers using the modified OPLS (Optimized Potentials for Liquid Simulations) force field with GROMACS in SPC/E water[44–47] (Figure S4-S6). Langevin dynamics simulations of the CG model with the HOOMD package[48] effectively reproduce the directional crystallization of P*i*PrOx (Figure 2B). Elongated ribbons with patches oriented parallel were obtained in simulations initialized from polymers with random starting conformations in agreement with the crystal structure of P*i*PrOx (Figure S7).

The fundamental unit of the targeted self-assembly process is a hybrid SiO$_2$ NP covalently grafted with P*i*PrOx chains on the NP surface. The SiO$_2$ NP is represented by a sphere of radius ten times the radius of one CG polymer bead (6.4 Å). P*i*PrOx chains are grafted with harmonic bonds at random locations on the spherical surface at a density comparable to the grafting density determined in the experiment (Table 2). We use 0.35 chains/nm$^2$, which corresponds to 45 chains per NP (Figure 2C and Figure S8).

**Table 3. Size and Cloud Point Characterization of the Synthesized P*i*PrOx and P*i*PrOx@SiO$_2$.**

| Sample | | | [a] T$_{CP}$ |
|---|---|---|---|
| P*i*PrOx$_{30}$ | | | 43 |
| P*i*PrOx$_{100}$ | | | 39 |
| [b] σ (P*i*PrOx$_{100}$@SiO$_2$) | [c] R$_h$ | [d] R$_h$ | [a] T$_{cp}$ |
| 0.11 | 15 | 18 | - |
| 0.18 | 15 | 18 | - |
| 0.35 | 15 | 23 | 45 |
| 0.47 | 15 | 24 | 42 |
| 0.24 | 25 | 36 | 44 |
| [b] σ (P*i*PrOx$_{30}$@SiO$_2$) | [c] Rh | [d] Rh | [a] Tcp |
| 0.70 | 24 | 28 | 47 |

[a] Cloud point temperature T$_{CP}$ of free polymers and the core-shell hybrid NPs in °C as determined by UV/vis measurement. [b] Grafting density in chains/nm$^2$. [c] Hydrodynamic radii of the SiO$_2$ core in nm as determined by DLS. [d] Hydrodynamic radii of the P*i*PrOx@SiO$_2$ core-shell NPs in nm as determined by DLS.

## Results and Discussion

### Solution Characterization of P*i*PrOx@SiO$_2$

Free P*i*PrOx polymers exhibit a lower critical solution temperature (LCST) in water of about 36 °C, depending on the presence of hydrophilic or hydrophobic end-groups and also the molecular weight.[38,49] We were interested in whether this LCST can be exploited for the self-assembly of P*i*PrOx@SiO$_2$ core-shell hybrid NPs. Preliminary observations indicated that P*i*PrOx@SiO$_2$ in water form a stable dispersion at room temperature. But it turns cloudy upon heating depending on the grafting density of P*i*PrOx chains on the SiO$_2$ surface, turns cloudy upon heating. We studied the thermosensitive behavior of P*i*PrOx and P*i*PrOx@SiO$_2$ in detail by measuring cloud point temperatures (T$_{CP}$) through UV–visible spectroscopy (UV–vis) of aqueous P*i*PrOx solutions and P*i*PrOx@SiO$_2$ dispersions (1.0 wt %). Samples were heated at a constant heating rate (0.2 °C/min) in a temperature range between 25 °C and 72 °C. T$_{CP}$ was defined as the temperature where the transmittance progressively decreased to 50%.[49]

First, we analyzed the effect of the chain length of free P*i*PrOx polymers on T$_{CP}$ (Table 3). T$_{CP}$ shifts to lower temperature with DP increasing from 30 to 100 in accordance with literature data.[50] We also investigated the influence of grafting P*i*PrOx on SiO$_2$ NPs, whereby the grafting density was varied from σ = 0.11 to 0.47 chain/nm$^2$ (Table 3). No phase transition was observed for σ below 0.18 chain/nm$^2$. However, when σ increased from 0.35 to 0.47 chain/nm$^2$, the observed T$_{CP}$ decreased by about 3 °C. Our data thus demonstrate that T$_{CP}$ can be adjusted by varying grafting density and increases with increasing overall hydrophilicity.

### P*i*PrOx Crystallization

Prolonged heating of P*i*PrOx above the cloud point temperature induces crystallization in dilute aqueous solution by the formation of crystalline nanofibers *via* directional crystallization.[35,51] To test how solution crystallization of P*i*PrOx is influenced when P*i*PrOx chains are covalently grafted onto SiO$_2$ NPs, we characterized the crystallization process of P*i*PrOx@SiO$_2$ by X-ray diffraction (XRD) and *in situ* ATR.

We recorded XRD patterns for samples formed upon incubation of P*i*PrOx$_{100}$@SiO$_2$ solutions (0.1 mg/mL and σ = 0.35 chain/nm$^2$) at 65 °C for various time periods. After incubation, the samples were cooled and freeze-dried. The solid residues were recovered and placed on a sample plate. XRD traces after 0, 6, and 28 hours incubation periods are presented in Figure 3A. While P*i*PrOx$_{100}$@SiO$_2$ before heating reveals mostly an amorphous pattern, the XRD trace recorded for a P*i*PrOx$_{100}$@SiO$_2$ sample heated for 6 hours already develops two diffraction peaks at 2Θ = 8.15 and 24 degrees, which can be assigned to semicrystalline P*i*PrOx and increases in intensity after 28 hours.[35]

We characterized the molecular interactions during crystallization by *in situ* IR spectroscopy. We focused on the C=O stretching band (1680 to 1570 cm$^{-1}$), which is well-studied in the literature.[52,53] Temperature- and time-dependent ATR IR measurements were performed on P*i*PrOx$_{100}$@SiO$_2$ (σ = 0.35 chain/nm$^2$) in D$_2$O solution (c = 5 mg/mL) kept sealed between the ATR surface and a cap to prevent solvent evaporation during the measurement (Figure S9A).

IR spectra were collected during a heating cycle. As shown by arrows in Figure 3D for the beginning temperature of 25 °C and the end temperature of 65 °C, the C=O stretching band shifts to larger wave numbers, and spectral intensity decrease. A systematic analysis during the heating cycles with intervals of 5 °C reveals a gradual peak shift from 1605 to 1625 cm$^{-1}$ during the heating cycle (Figure S9B). This peak shift towards higher wavenumbers indicates that interactions between carbonyl groups and water molecules gradually break during the formation of crosslinking hydrogen bond bridges between P$i$PrOx chains and free C=O.[54,55]

Two stages, nucleation, and crystal growth, were distinguished for P$i$PrOx crystallization.[53] Nucleation can be induced by annealing the solution above the LCST, and growth is critical for ligand chain ordering. We examined the growth stage by time-dependent characterization of the C=O stretching bands. ATR spectra were collected during 48 hours of annealing at 65 °C (Figure 3C, D, and Figure S9C). The C=O stretching band gradually shifts from 1625 to 1640 cm$^{-1}$ over 20 hours of heating, indicative of a dehydration process during crystallization.[52]

For more in-depth analysis, we plotted the second derivatives of the IR spectra (Figure 3C). There are three peaks at 1644, 1630, and 1610 cm$^{-1}$. Upon annealing, the peaks at 1630 and 1610 cm$^{-1}$ gradually disappear while the peak at 1644 cm$^{-1}$ appears. The former two peaks are assigned to the stretching bands of C=O in C=O---D–O–D---O=C hydrogen bonds and C=O---D$_2$O hydrogen bonds, respectively, while the latter one arises from ordered C=O in crystalline P$i$PrOx chains.[52,55] These spectral changes of the C=O stretching band indicate that the crystallization of P$i$PrOx at 65 °C leads to the cleavage of C=O related hydrogen bonds and an ordered assembly of carbonyl groups. We summarize our findings in this section with a schematic illustration in Figure 3D. P$i$PrOx crystallization in an aqueous solution proceeds in two steps: dehydration followed by crystallization.

### Solution behavior of P$i$PrOx@SiO$_2$ in water

The solution behavior and stability of the P$i$PrOx@SiO$_2$ hybrid materials in an aqueous solution are governed by the basic structural characteristics of the grafted polymers and the NP cores. These include molecular weight, dispersity, chemical composition and grafting density of P$i$PrOx, as well as size, geometry, and morphology for the core NP.[18,20,56] These parameters can significantly control the solution stability of polymer-grafted NP initially and during any heating protocol. We therefore further investigated the effects of grafting density, core size, annealing temperature, and overall concentration on the solution stability and behavior of P$i$PrOx@SiO$_2$ upon heating in a dilute aqueous solution.

The NP concentration represents a critical parameter governing the evolution of the spatial arrangement of NPs. Our visual observation indicates that the concentrations in the range of 1 to 10 mg mL$^{-1}$ resulted in the rapid formation of a macroscopic precipitate during the heating. Indeed, samples with high concentrations showed white microscopical precipitation nearly during a maximum of one hour of heating. This phenomenon is due to probably the fast co-aggregation of core-shell NPs during solution heating. We, therefore, decreased the concentration to values between 0.01 and 0.5 mg mL$^{-1}$ for all following investigations to prevent premature sedimentation.

The effective NP interactions are largely dependent on the conformation of the polymer brushes on the NP surfaces, which in turn are affected by grafting density as well as by chain length.[57] We synthesized P$i$PrOx@SiO$_2$ hybrids with four different σ from 0.11 to 0.47 chain/nm$^2$ (Figure 4) for SiO$_2$ core radii of 15 nm and P$i$PrOx with a DP of 100 as exemplary samples. During heating, we monitored solution stability through DLS and UV–visible spectroscopy (UV–vis) at room temperature and 65 °C. In the latter case, the samples were heated at a constant temperature of 65 °C, and transmittance values were collected at different time intervals.

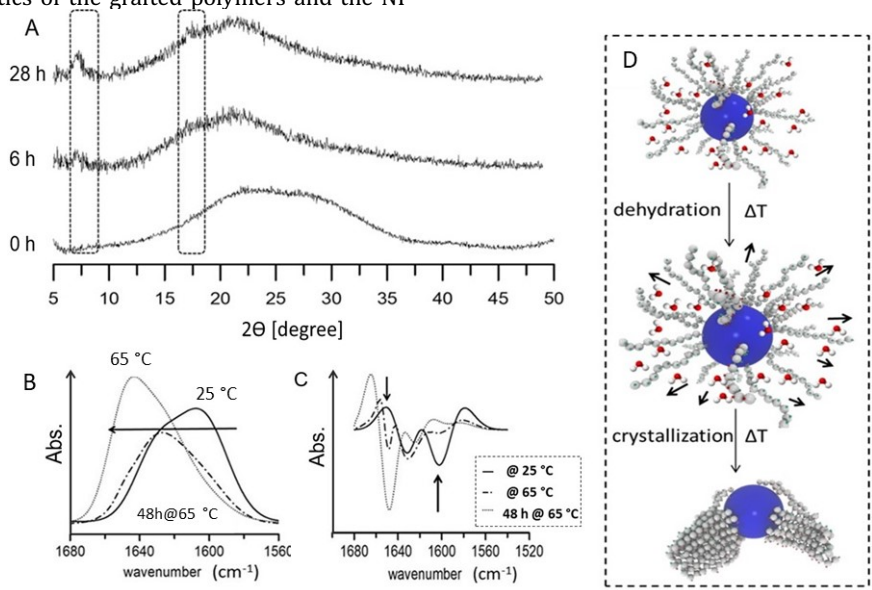

**Figure 3.** (A): XRD diffractograms of freeze-dried P$i$PrOx@SiO$_2$ (5 mg/mL, 0.35 chain/nm$^2$) after annealing a solution for 0, 6, and 28 hours. (B): ATR spectra in the carbonyl region (C) and second derivatives. (D): Schematic depiction of the P$i$PrOx@SiO$_2$ dehydration, followed by crystallization.

Both visual observation and DLS measurements indicate that all samples form stable dispersions at room temperature without signs of sedimentation even after prolonged storage of up to one month. We also determined the solution stability of P*i*PrOx$_{100}$ and P*i*PrOx$_{100}$@SiO$_2$ hybrids at 65 °C (Figure 4). Transmittance in all cases except for 0.11 chains/nm$^2$ started to increase, presumably due to some precipitate formation. Comparing the evolution of transmittance (or absorption) over time, sedimentation probability follows the order: 0.18 chain/nm$^2$ > P*i*PrOx$_{100}$ > 0.35 chain/nm$^2$ > 0.47 chain/nm$^2$ > 0.11 chain/nm$^2$. P*i*PrOx itself shows macroscopic precipitation after about 4 to 5 hours of heating. On the other hand, P*i*PrOx$_{100}$@SiO$_2$ samples seem to reveal an improved stability against sedimentation and by increasing σ, this stability appears to be more pronounced. We can tentatively explain the effect of steric stabilization, thereby preventing agglomeration due to attractive van der Waals and depletion forces. Samples with a grafting density of 0.11 chain/nm$^2$ did not exhibit signs of precipitation, which is in line with the lack of temperature sensitivity discussed above.

### Crystallization-Driven Self-assembly of P*i*PrOx@SiO$_2$

P*i*PrOx can undergo directional crystallization upon prolonged heating above the cloud point temperature in aqueous media. We have demonstrated this behavior in previous work for the formation of anisotropic superstructures in block copolymers.[38,39] Here, we employ isotropic and spherical P*i*PrOx@SiO$_2$ as building blocks. Although the directional crystallization of P*i*PrOx has been reported before,[35,58] the effect of covalent anchorage of P*i*PrOx chains to NP surfaces has not been investigated. One obvious reason is that NPs often undergo secondary aggregation. To deal with this challenge, we investigate factors that control the solution behavior of P*i*PrOx@SiO$_2$ core-shell hybrids during crystallization and heating.

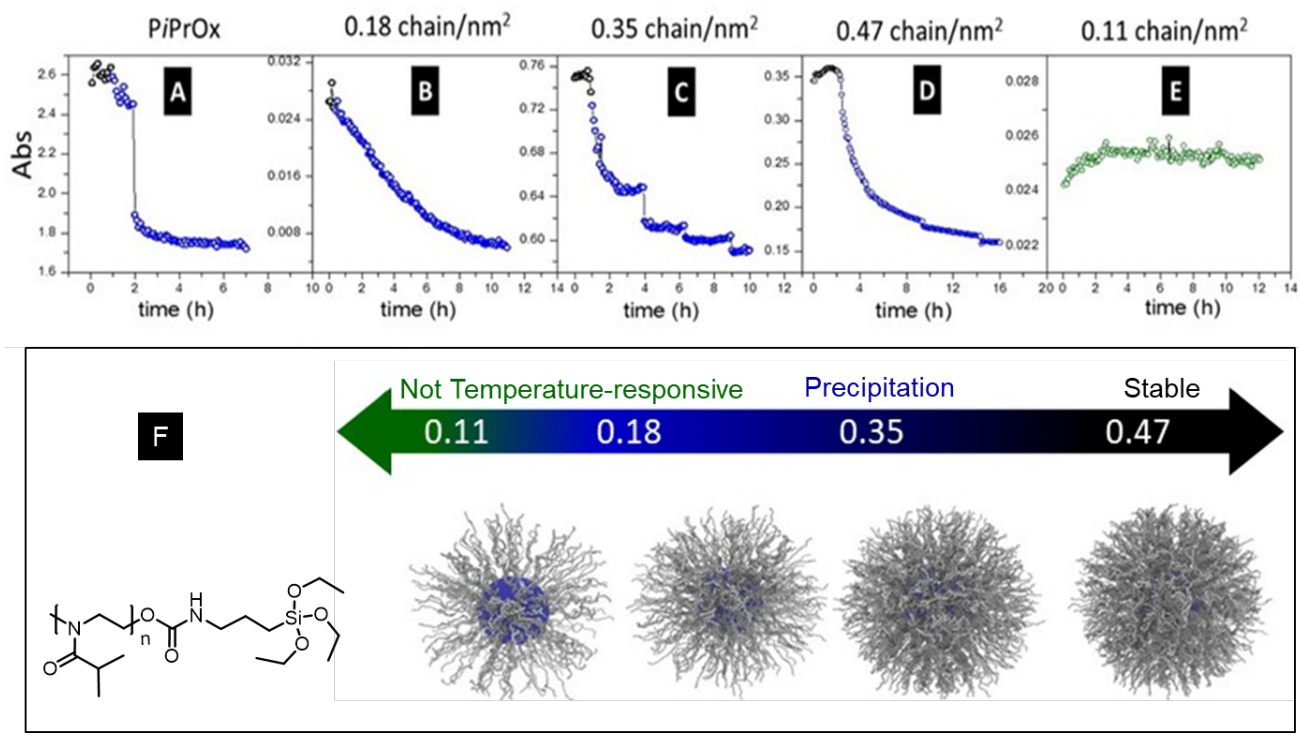

**Figure 4.** (A-E): Time-dependent solution stability of P*i*PrOx and P*i*PrOx@SiO$_2$ (DP: 100 and R$_h$ 23 nm) with different grafting densities upon prolonged heating (65° C). (F): Schematic illustration of the relation between grafting density and solution behavior of P*i*PrOx and P*i*PrOx@SiO$_2$. Samples show three different types of behavior, which are illustrated by black, blue, and green colors. Black color indicates turbid but colloidally stable dispersions, blue color depicts samples that slowly agglomerate and/or precipitate, and green shows samples without significant temperature response in our experiments.

The aqueous P*i*PrOx$_{100}$@SiO$_2$ solutions were investigated by heating from 25 to 70 °C and also prolonged heating for ~ 16 h at 65 °C was carried out. In all cases, the scattering intensity and the hydrodynamic size were monitored (Figures S10 and S11). As an exemplary case, both scattering intensity and time-dependent size for P*i*PrOx@SiO$_2$ solutions with 0.1 mg/mL and σ 0.35 chains/nm$^2$ are illustrated in Figure S10, A-B. Thereby, R$_h$ of P*i*PrOx@SiO$_2$ decreases during heating from 25 to 70 °C, whereas the scattering intensity increases up to ~ 45 °C, followed by a sharp drop. This might indicate the formation of secondary aggregates with larger size distribution.

Along the same line, sample solutions that are kept for a longer time at 65 °C also show increasing hydrodynamic radii until more than one mode is present, leading to the assumption that two different aggregate populations coexist (Figure S10C). In the case of P*i*PrOx$_{100}$@SiO$_2$ with σ = 0.47 chain/nm$^2$, increasing size after 1-hour heating could be observed (Figure S11). On the other hand, P*i*PrOx$_{100}$@SiO$_2$ with σ = 0.11 and 0.18 chains/nm$^2$ exhibited a sharp drop in hydrodynamic size after 2 h, which hints towards fast sedimentation, presumably due to an insufficient amount of P*i*PrOx present on the NP surface.

According to DLS results, there seems to be a clear difference regarding heating-induced aggregation in dependence of the grafting density and, hence, samples annealed for 24 h at 65 °C were also investigated *via* transmission electron microscopy (TEM) and scanning electron microscopy (SEM).

In the case of the lower grafting densities (0.11 and 0.18 chain/nm$^2$), rather ill-defined aggregates as well as individual particles are found (Figure S12). Crystallization of the P*i*PrOx shell leads to some agglomerates as well as single particles as a macroscopic precipitate.

The time-dependent DLS data for higher grafting densities suggests further aggregation upon heating, yet leading to rather stable dispersions (Figure S11). Figure 5 (and Figure S16-S19) illustrate the morphology of these aggregates after drop-casting in TEM and SEM images for P*i*PrOx$_{100}$@SiO$_2$ as a representative sample with σ = 0.35 chain/nm$^2$ at different concentrations. As can be seen, surprisingly rather long fibers decorated with SiO$_2$ NP could be observed, which are several micrometers in length and connected to node centers with cross-sectional diameters of several μm (Figure 5 and Figure S12-S15). Besides that, we also observed some individual SiO$_2$ NP and free fibers decorated by SiO$_2$ in TEM and SEM images (Figure 5). These structures were found with diameters of around 10 nm and lengths of 0.5 to 4 μm. These superstructures were investigated by TEM and SEM and were found to match the fibers described in our previous study for block copolymers.[38,39] These aggregates can be observed also in P*i*PrOx$_{100}$@SiO$_2$ with higher σ (0.47 chain/nm$^2$, Figure S13).

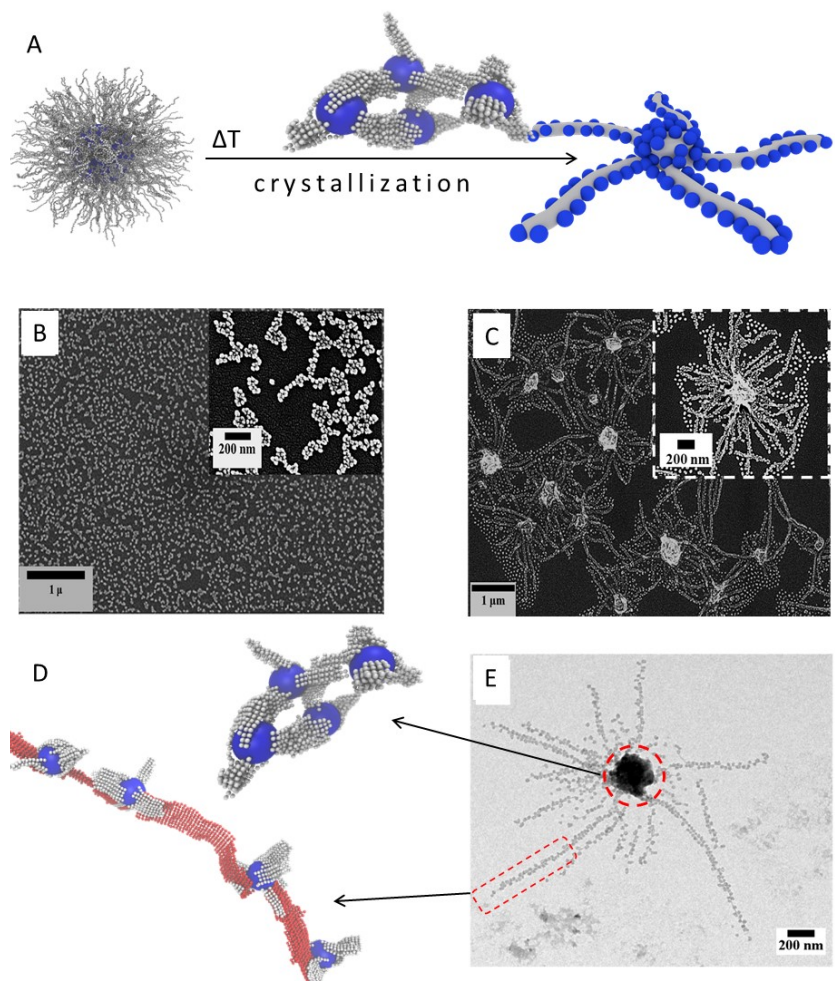

Figure 5 (A): Schematic depiction of superstructure formation during crystallization of P*i*PrOx in P*i*PrOx@SiO$_2$. (B): SEM images of P*i*PrOx$_{100}$@SiO$_2$ (0.1 mg/mL, 0.35 chain/nm$^2$). (C and E): SEM and TEM images of P*i*PrOx$_{100}$@SiO$_2$ (0.1 mg/mL, 0.35 chain/nm$^2$) solution after 24 hours heating at 65°C. (D): Structures observed in molecular dynamics simulations. The grafted polymers are represented by white and unbound polymers are represented by red.

To understand the interactions governing the self-assembly of P*i*PrOx@SiO$_2$ NPs, we also performed molecular dynamics simulations with a grafting density of 0.35 chain/nm$^2$ comparable with the experiment. In our simulations, we considered four P*i*PrOx@SiO$_2$ particles in the simulation box as a starting point while the real experimental system contains a large number of such particles. We observed aggregation of NP mediated by the grafted polymers (Figure 5D and S21). Extrapolating this result to a system with a large number of P*i*PrOx@SiO$_2$, we expect the formation of a quasi-bulk three-dimensional network, supporting the formation of central nodes as observed in SEM analysis. Nevertheless, the formation of the diverging arms decorated with SiO$_2$ NP cannot be explained using this mechanism, hinting towards another self-assembly mechanism being involved as well. The possibility of multiple mechanisms also supported by the presence of free arms consisting of long fibers without attached SiO$_2$ NP (Figure 5E and S19). This led us to speculate that prolonged heating leads to the formation of nanoribbons as reported as a characteristic property of P*i*PrOx and the ribbons can then interact with P*i*PrOx@SiO$_2$ NPs to form the arms diverging from the central node which are decorated with NPs (Figure S20 and S21). The formation of ribbons was probably initiated by crystallization of polymer chains that are physically adsorbed in the entangled polymer matrix on the NP surface in the beginning and eventually released at elevated temperatures (Figure S22). These ribbons may be further elongated by the crystallization driven detachment of polymers from the surface.

To verify the possibility of formation of P*i*PrOx@SiO$_2$ arms *via* P*i*PrOx nanoribbons, we set up simulations with unbound polymer ribbons and P*i*PrOx@SiO$_2$ particles. We observed the interaction of the polymers grafted on the NPs with the nanoribbons (Figure S21, D and H). In this scenario, the long arms are formed by nanoribbons interconnected *via* P*i*PrOx@SiO$_2$ particles. We observe that grafted polymers interact with ribbons perpendicular to the amide interaction and contribute to the thickening of the ribbon (Figure S20 D, and E).

From the microscopic images and MD simulations, we noted that the P*i*PrOx@SiO$_2$ particles and nanoribbons interact in two ways (Figure S20): 1) The P*i*PrOx@SiO$_2$ particles interact with the surface of long nanoribbons using hydrophobic interactions and 2) one P*i*PrOx@SiO$_2$ particle interacts simultaneously with two ribbons, possibly even stitching both together. Combinations of both interactions then lead to the formation of long ribbons, where a maximum of two NPs are present at the sides (Figure 5).

Observations from molecular dynamics simulations support the formation of structures by P*i*PrOx@SiO$_2$ *via* two self-assembly mechanisms. First, central nodes are formed by the interaction between NPs through P*i*PrOx crystallization upon prolonged heating, and the process is facilitated by the presence of some unbound polymer chains. Second, arms are formed by the interaction of P*i*PrOx@SiO$_2$ particles with nanoribbons formed from free P*i*PrOx and individual P*i*PrOx@SiO$_2$.

Another possibility for the formation of free nanoribbons in the absence of free polymers would be the detachment of a small fraction of polymeric ligands during the heating process. After that, a similar mechanism as discussed above would be in place. However, due to the covalent attachment of the P*i*PrOx chains in our set of samples, this scenario is rather unlikely and we speculate that we indeed observe the effect of a small fraction of physically bound polymeric chains being present. Nevertheless, all efforts to prove the presence of free P*i*PrOx in these samples were unsuccessful so far. Besides, self-assembly of P*i*PrOx@SiO$_2$ (0.35 chain/nm$^2$, 0.5 mg/mL, at 65 °C) was observed in the presence of the different amounts of free P*i*PrOx with concentrations of 0.01 mg/mL, 0.1, and 0.5 mg/mL (*e.g.*, ratio 1:1 of P*i*PrOx to P*i*PrOx@SiO$_2$). In this case, SEM micrographs indicate mainly the formation of microspheres similar to the structures reported by Schlaad and coworkers for P*i*PrOx homopolymers (Figure S23).[36]

Core and shell size are expected to have a significant impact on solution stability, thus giving a handle to control the morphology of P*i*PrOx@SiO$_2$ aggregates. The influence of the core size was investigated by comparing the aggregate morphology for different core sizes, here being 15 nm and 25 nm in radii. Also here, the obtained particles were evaluated with regard to grafting density and size with DLS, TGA, and TEM (Figures S5, S11, and S14). Afterward, P*i*PrOx@SiO$_2$ with a core size of 25 nm in radii was annealed under comparable conditions and the morphology was evaluated with SEM (Figure S14). In this case, the central nodes are not formed but only the arms decorated by SiO$_2$ are present, which we attribute to altered solution stability. In addition, the influence of the shell thickness was also controlled during the crystallization. Comparing data, we observe that shorter P*i*PrOx chains with DP of 30 seem not able to form decorated nanoribbons, and mostly ill-defined aggregates are found (Figure S15).

Another important parameter is the temperature at which the sample solutions are annealed - we therefore also investigated prolonged heating for P*i*PrOx$_{100}$@SiO$_2$ with σ = 0.35 chains/nm$^2$ at a concentration of 0.1 mg/mL at 40, 50, 60, and 70 °C respectively (Figure S16). While the obtained superstructures at 50, 60, and 70 °C show the formation of tentacles decorated with SiO$_2$ NP and node centers, P*i*PrOx$_{100}$@SiO$_2$ heated at 40 °C shows only spherical aggregates with an average size of ~ 200 nm (Figure S16). The observed spherical structures are formed mostly by aggregation of several small P*i*PrOx$_{100}$@SiO$_2$ NP. These observations, in our opinion, support the assumption of different crystallization mechanisms and highlight the importance of concentration and temperature for the initial formation of well-defined anisotropic superstructures.

To support both the fusion mechanism of NP and nanoribbon formation during the crystallization, the evolution of the morphology during the heating and crystallization process was investigated by observing the morphology at different times during heating. Thereby, Figure S17 exhibits the SEM images of P*i*PrOx@SiO$_2$ (0.35 chain/nm$^2$, 0.1 mg/mL, at 65 °C) which are collected and drop-cast on SEM grids after 2, 4, 8, 12, and 24 hours of heating. As our data shows, samples show single particles without any superstructure formation before heating (Figure 5A). After 2 hours the morphology changed considerably. SEM shows some clusters of aggregates consisting of several hundred single NPs and short fibers with attached SiO$_2$ particles simultaneously (Figure S17). However, short tentacles wrap the structures in many cases. With increasing time of annealing and crystallinity of the sample, the number of clusters increases, and longer anisotropic superstructures (arms) are formed. Eventually, as observed in the sample taken at 24 h, crystallization resulted in the formation of long tentacles with a lower overall amount of cluster nodes.

To further observe the structural evolution, confocal laser scanning microscopy (CLSM) was used as an online technique. Hereby, an aqueous solution of P*i*PrOx$_{100}$@SiO$_2$ (0.35 chain/nm$^2$, 1 wt %) was placed between two sealed microscope slides, which are shielded by silicon tape. The changes in the appearance of a specific area of the specimen were recorded during heating the solution for several hours (Figure S24 and Supporting information 2 (video)). Our observation indicates that different phases are formed after 1 to 2 hours of heating which appeared as shadows. At the same time, darker spots appear which probably indicate the formation of larger superstructures. Shadows with the shape of chains start to appear after 4 hours of heating, merging into larger structures with increasing time. We think that this qualitatively supports the two-step aggregation mechanism suggested by a combination of simulation and experiments.

## Conclusion

We open a bottom-up strategy for the self-assembly of isotropic and spherical core–shell hybrids to an anisotropic superstructure of organic/inorganic hybrid materials by directional crystallization of the poly(2-*iso*-propyl-2-oxazoline) (P*i*PrOx) in aqueous solution. For this purpose, we grafted the polymeric shell isotropically by end modification of the polymer shell. In the next step, the directional crystallization of PiPrOx in water was investigated by heating the solution for 24 h to 65 °C. The crystallization was monitored in situ by DLS, while the resulting structures were investigated in detail by electron microscopy (TEM and SEM). Several different assembly structures were identified for P*i*PrOx@SiO$_2$, presumably owing to differences in the hydrophobic/hydrophilic interactions and solution stability of core-shell hybrids respectively. While core-shell hybrids with lower grafting density and larger size of the core formed short arms of the fibers decorated with SiO$_2$, the higher grafting density with a smaller core illustrates a tentacle decorated with SiO$_2$ NPs with several micrometers in length and the center of the node with cross-sectional diameters of several micrometers. These first results toward the formation of anisotropic hybrid materials by core-shell structures with crystalline shells demonstrate the potential of such materials for the controlled and stepwise directional assembly in solution into 2D or 3D structures. Optimization of shell composition will be the subject of further investigations.

## ASSOCIATED CONTENT

Experimental details for the synthesis of core-shell hybrids, additional experimental results for determination of particle size, crystallization, cloud points, and crystallization mechanism, additional SEM and TEM micrographs of aggregates for crystallization and after thermal treatment, confocal laser scanning microscopy images of P*i*PrOx@SiO$_2$, and schematic illustration of nano ribbon formation mechanism.

## AUTHOR INFORMATION

### Corresponding Authors

#### Felix H. Schacher

Friedrich-Schiller University Jena, Lessingstraße 8, D-07743 Jena, Germany
Email: felix.schacher@uni-jena.de

#### Michael Engel

Friedrich-Alexander-Universität Erlangen-Nürnberg, Cauerstrasse 4, D-91058 Erlangen, Germany.
Email: michael.engel@fau.de

### Author Contributions

The manuscript was written through contributions of all authors. A.N. performed all experimental investigation and characterization. A.M. performed all simulations. A.M. and F.T. developed the coarse-grained model and analyzed simulation data. M.E. and F.H. supervised the project. F.H. is responsible for the project concept.

### Funding Sources

A.M., F.T., and M.E. acknowledge support by the Deutsche Forschungsgemeinschaft (DFG, German Research Foundation) Project-ID 416229255—SFB 1411. This project was further funded by the Deutsche Forschungsgemeinschaft (Projects B02, B05 and Z02 within the Sonderforschungsbereich SFB/TRR 234 "Catalight", project ID: 364549901)

### Notes

Any additional relevant notes should be placed here.

## ACKNOWLEDGMENT

The authors thank the TEM facilities of the Jena Center for Soft Matter (JCSM) were established with a grant from the German Research Council (DFG) and the European Fund for Regional Development (EFRE). We further thank Yves Carstensen, Dr. Johannes Buchheim, and Dr. Philip Biehl for SEM, TEM, and XRD analysis respectively. Computational resources and support provided by the Erlangen Regional Computing Center (RRZE) are gratefully acknowledged.

## ABBREVIATIONS

## REFERENCES

(1) Nie, Z.; Petukhova, A.; Kumacheva, E. Properties and Emerging Applications of Self-Assembled Structures Made from Inorganic Nanoparticles. *Nat. Nanotechnol.* **2010**, *5* (1), 15.

(2) Yi, C.; Yang, Y.; Liu, B.; He, J.; Nie, Z. Polymer-Guided Assembly of Inorganic Nanoparticles. *Chem. Soc. Rev.* **2020**, *49* (2), 465–508.

(3) Zhang, Z.; Yang, P.; Xu, H.; Zheng, H. Surface Enhanced Fluorescence and Raman Scattering by Gold Nanoparticle Dimers and Trimers. *J. Appl. Phys.* **2013**, *113* (3), 33102.

(4) Lee, A.; Andrade, G. F. S.; Ahmed, A.; Souza, M. L.; Coombs, N.; Tumarkin, E.; Liu, K.; Gordon, R.; Brolo, A. G.; Kumacheva, E. Probing Dynamic Generation of Hot-Spots in Self-Assembled Chains of Gold Nanorods by Surface-Enhanced Raman Scattering. *J. Am. Chem. Soc.* **2011**, *133* (19), 7563–7570.

(5) Bian, Z.; Tachikawa, T.; Zhang, P.; Fujitsuka, M.; Majima, T. Au/TiO2 Superstructure-Based Plasmonic Photocatalysts Exhibiting Efficient Charge Separation and Unprecedented Activity. *J. Am. Chem. Soc.* **2014**, *136* (1), 458–465.

(6) Tao, H.; Galati, E.; Kumacheva, E. Temperature-Responsive Self-Assembly of Nanoparticles Grafted with UCST Polymer Ligands. *Macromolecules* **2018**, *51* (15), 6021–6027.

(7) Park, W. M.; Champion, J. A. Colloidal Assembly of Hierarchically Structured Porous Supraparticles from Flower-Shaped Protein–Inorganic Hybrid Nanoparticles. *ACS Nano* **2016**, *10* (9), 8271–8280.

(8) Rey, M.; Law, A. D.; Buzza, D. M. A.; Vogel, N. Anisotropic Self-Assembly from Isotropic Colloidal Building Blocks. *J. Am. Chem. Soc.* **2017**, *139* (48), 17464–17473.

(9) Min, Y.; Akbulut, M.; Kristiansen, K.; Golan, Y.; Israelachvili, J. The Role of Interparticle and External Forces in Nanoparticle Assembly. In *Nanoscience and Technology: A Collection of Reviews from Nature Journals*; World Scientific, 2010; pp 38–49.

(10) Singh, G.; Chan, H.; Baskin, A.; Gelman, E.; Repnin, N.; Král, P.; Klajn, R. Self-Assembly of Magnetite Nanocubes into Helical Superstructures. *Science (80-. ).* **2014**, *345* (6201), 1149–1153.

(11) Correa-Duarte, M. A.; Pérez-Juste, J.; Sánchez-Iglesias, A.; Giersig, M.; Liz-Marzán, L. M. Aligning Au Nanorods by Using Carbon Nanotubes as Templates. *Angew. Chemie* **2005**, *117* (28), 4449–4452.

(12) Park, J. Il; Nguyen, T. D.; de Queirós Silveira, G.; Bahng, J. H.; Srivastava, S.; Zhao, G.; Sun, K.; Zhang, P.; Glotzer, S. C.; Kotov, N. A. Terminal Supraparticle Assemblies from Similarly Charged Protein Molecules and Nanoparticles. *Nat. Commun.* **2014**, *5* (1), 1–9.

(13) Vaia, R. A.; Maguire, J. F. Polymer Nanocomposites with Prescribed Morphology: Going beyond Nanoparticle-Filled Polymers. *Chem. Mater.* **2007**, *19* (11), 2736–2751.

(14) Gröschel, A. H.; Löbling, T. I.; Petrov, P. D.; Müllner, M.; Kuttner, C.; Wieberger, F.; Müller, A. H. E. Janus Micelles as Effective Supracolloidal Dispersants for Carbon Nanotubes. *Angew. Chemie Int. Ed.* **2013**, *52* (13), 3602–3606.

(15) Vogel, N.; Retsch, M.; Fustin, C.-A.; del Campo, A.; Jonas, U. Advances in Colloidal Assembly: The Design of Structure and Hierarchy in Two and Three Dimensions. *Chem. Rev.* **2015**, *115* (13), 6265–6311.

(16) Liu, K.; Nie, Z.; Zhao, N.; Li, W.; Rubinstein, M.; Kumacheva, E. Step-Growth Polymerization of Inorganic Nanoparticles. *Science (80-. ).* **2010**, *329* (5988), 197–200.

(17) Song, J.; Cheng, L.; Liu, A.; Yin, J.; Kuang, M.; Duan, H. Plasmonic Vesicles of Amphiphilic Gold Nanocrystals: Self-Assembly and External-Stimuli-Triggered Destruction. *J. Am. Chem. Soc.* **2011**, *133* (28), 10760–10763.

(18) Yi, C.; Zhang, S.; Webb, K. T.; Nie, Z. Anisotropic Self-Assembly of Hairy Inorganic Nanoparticles. *Acc. Chem. Res.* **2017**, *50* (1), 12–21. https://doi.org/10.1021/acs.accounts.6b00343.

(19) Akcora, P.; Liu, H.; Kumar, S. K.; Moll, J.; Li, Y.; Benicewicz, B. C.; Schadler, L. S.; Acehan, D.; Panagiotopoulos, A. Z.; Pryamitsyn, V. Anisotropic Self-Assembly of Spherical Polymer-Grafted Nanoparticles. *Nat. Mater.* **2009**, *8* (4), 354–359.

(20) Chancellor, A. J.; Seymour, B. T.; Zhao, B. Characterizing Polymer-Grafted Nanoparticles: From Basic Defining Parameters to Behavior in Solvents and Self-Assembled Structures. ACS Publications 2019.

(21) Guo, Y.; Harirchian-Saei, S.; Izumi, C. M. S.; Moffitt, M. G. Block Copolymer Mimetic Self-Assembly of Inorganic Nanoparticles. *ACS Nano* **2011**, *5* (4), 3309–3318.

(22) Wang, B.; Li, B.; Dong, B.; Zhao, B.; Li, C. Y. Homo-and Hetero-Particle Clusters Formed by Janus Nanoparticles with Bicompartment Polymer Brushes. *Macromolecules* **2010**, *43* (22), 9234–9238.

(23) He, J.; Liu, Y.; Babu, T.; Wei, Z.; Nie, Z. Self-Assembly of Inorganic Nanoparticle Vesicles and Tubules Driven by Tethered Linear Block Copolymers. *J. Am. Chem. Soc.* **2012**, *134* (28), 11342–11345.

(24) Glotzer, S. C.; Solomon, M. J. Anisotropy of Building Blocks and Their Assembly into Complex Structures. *Nat. Mater.* **2007**, *6* (8), 557.

(25) Liu, Y.; Klement, M.; Wang, Y.; Zhong, Y.; Zhu, B.; Chen, J.; Engel, M.; Ye, X. Macromolecular Ligand Engineering for Programmable Nanoprism Assembly. *J. Am. Chem. Soc.* **2021**, *143* (39), 16163–16172.

(26) Li, C. Y. Anisotropy Unnecessary. *Nat. Mater.* **2009**, *8* (4), 249–250.

(27) Zhou, S.; Sakamoto, T.; Wang, J.; Sugawara-Narutaki, A.; Shimojima, A.; Okubo, T. One-Dimensional Assembly of Silica Nanospheres: Effects of Nonionic Block Copolymers. *Langmuir* **2012**, *28* (37), 13181–13188.

(28) Martin, T. B.; Seifpour, A.; Jayaraman, A. Assembly of Copolymer Functionalized Nanoparticles: A Monte Carlo Simulation Study. *Soft Matter* **2011**, *7* (13), 5952–5964.

(29) Wang, J.; Xia, H.; Zhang, Y.; Lu, H.; Kamat, R.; Dobrynin, A. V; Cheng, J.; Lin, Y. Nucleation-Controlled Polymerization of Nanoparticles into Supramolecular Structures. *J. Am. Chem. Soc.* **2013**, *135* (31), 11417–11420.

(30) He, W.-N.; Xu, J.-T. Crystallization Assisted Self-Assembly of Semicrystalline Block Copolymers. *Prog. Polym. Sci.* **2012**, *37* (10), 1350–1400.

(31) Mai, Y.; Eisenberg, A. Self-Assembly of Block Copolymers. *Chem. Soc. Rev.* **2012**, *41* (18), 5969–5985.

(32) Schmelz, J.; Schacher, F. H.; Schmalz, H. Cylindrical Crystalline-Core Micelles: Pushing the Limits of Solution Self-Assembly. *Soft Matter* **2013**, *9* (7), 2101–2107.

(33) Gilroy, J. B.; Gädt, T.; Whittell, G. R.; Chabanne, L.; Mitchels, J. M.; Richardson, R. M.; Winnik, M. A.; Manners, I. Monodisperse Cylindrical Micelles by Crystallization-Driven Living Self-Assembly. *Nat. Chem.* **2010**, *2* (7), 566–570.

(34) Wang, X.; Guerin, G.; Wang, H.; Wang, Y.; Manners, I.; Winnik, M. A. Cylindrical Block Copolymer Micelles and Co-Micelles of Controlled Length and Architecture. *Science (80-. ).* **2007**, *317* (5838), 644–647.

(35) Demirel, A. L.; Meyer, M.; Schlaad, H. Formation of Polyamide Nanofibers by Directional Crystallization in Aqueous Solution. *Angew. Chemie - Int. Ed.* **2007**, *46* (45), 8622–8624. https://doi.org/10.1002/anie.200703486.

(36) Diehl, C.; Černoch, P.; Zenke, I.; Runge, H.; Pitschke, R.; Hartmann, J.; Tiersch, B.; Schlaad, H. Mechanistic Study of the Phase Separation/Crystallization Process of Poly (2-Isopropyl-2-Oxazoline) in Hot Water. *Soft Matter* **2010**, *6* (16), 3784–3788.

(37) Giannini, C.; Ladisa, M.; Altamura, D.; Siliqi, D.; Sibillano, T.; De Caro, L.; Diehl, C.; Černoch, P.; Zenke, I.; Runge, H.; et al. Crystallisation-Driven Self-Assembly of Poly(2-Isopropyl-2-Oxazoline)-Block-Poly(2-Methyl-2-Oxazoline) above the LCST. *Soft Matter* **2015**, *6* (19), 87. https://doi.org/10.1021/acs.macromol.7b01639.

(38) Nabiyan, A.; Biehl, P.; Schacher, F. H. Crystallization vs Metal Chelation: Solution Self-Assembly of Dual Responsive Block Copolymers. *Macromolecules* **2020**, *53* (13), 5056–5067.

(39) Rudolph, T.; Lühe, M. von der; Hartlieb, M.; Norsic, S.; Schubert, U. S.; Boisson, C.; D'Agosto, F.; Schacher, F. H. Toward Anisotropic Hybrid Materials: Directional Crystallization of Amphiphilic Polyoxazoline-Based Triblock Terpolymers. *ACS Nano* **2015**, *9* (10), 10085–10098.

(40) Eckardt, O.; Pietsch, C.; Zumann, O.; von der Lühe, M.; Brauer, D. S.; Schacher, F. H. Well-Defined SiO2@ P (EtOx-stat-EI) Core–Shell Hybrid Nanoparticles via Sol-Gel Processes. *Macromol. Rapid Commun.* **2016**, *37* (4), 337–342.

(41) Stöber, W.; Fink, A.; Bohn, E. Controlled Growth of Monodisperse Silica Spheres in the Micron Size Range. *J. Colloid Interface Sci.* **1968**, *26* (1), 62–69.

(42) Wu, Z.; Beltran-Villegas, D. J.; Jayaraman, A. Development of a New Coarse-Grained Model to Simulate Assembly of Cellulose Chains Due to Hydrogen Bonding. *J. Chem. Theory Comput.* **2020**, *16* (7), 4599–4614.

(43) Starr, F. W.; Sciortino, F. Model for Assembly and Gelation of Four-Armed DNA Dendrimers. *J. Phys. Condens. Matter* **2006**, *18* (26), L347.

(44) Jorgensen, W. L.; Maxwell, D. S.; Tirado-Rives, J. Development and Testing of the OPLS All-Atom Force Field on Conformational Energetics and Properties of Organic Liquids. *J. Am. Chem. Soc.* **1996**, *118* (45), 11225–11236.

(45) Kaminski, G. A.; Friesner, R. A.; Tirado-Rives, J.; Jorgensen, W. L. Evaluation and Reparametrization of the OPLS-AA Force Field for Proteins via Comparison with Accurate Quantum Chemical Calculations on Peptides. *J. Phys. Chem. B* **2001**, *105* (28), 6474–6487.

(46) Berendsen, H. J. C.; van der Spoel, D.; van Drunen, R. GROMACS: A Message-Passing Parallel Molecular Dynamics Implementation. *Comput. Phys. Commun.* **1995**, *91* (1–3), 43–56.

(47) Berendsen, H. J. C.; Grigera, J. R.; Straatsma, T. P. The Missing Term in Effective Pair Potentials. *J. Phys. Chem.* **1987**, *91* (24), 6269–6271.

(48) Anderson, J. A.; Glaser, J.; Glotzer, S. C. HOOMD-Blue: A Python Package for High-Performance Molecular Dynamics and Hard Particle Monte Carlo Simulations. *Comput. Mater. Sci.* **2020**, *173*, 109363.

(49) Park, J.-S.; Akiyama, Y.; Winnik, F. M.; Kataoka, K. Versatile Synthesis of End-Functionalized Thermosensitive Poly (2-Isopropyl-2-Oxazolines). *Macromolecules* **2004**, *37* (18), 6786–6792.

(50) Katsumoto, Y.; Tsuchiizu, A.; Qiu, X.; Winnik, F. M. Dissecting the Mechanism of the Heat-Induced Phase Separation and Crystallization of Poly (2-Isopropyl-2-Oxazoline) in Water through Vibrational Spectroscopy and Molecular Orbital Calculations. *Macromolecules* **2012**, *45* (8), 3531–3541.

(51) Meyer, M.; Schlaad, H. Poly (2-Isopropyl-2-Oxazoline)– Poly (l-Glutamate) Block Copolymers through Ammonium-Mediated NCA Polymerization. *Macromolecules* **2006**, *39* (11), 3967–3970.

(52) Li, T.; Tang, H.; Wu, P. Molecular Evolution of Poly(2-Isopropyl-2-Oxazoline) Aqueous Solution during the Liquid-Liquid Phase Separation and Phase Transition Process. *Langmuir* **2015**, *31* (24), 6870–6878.

(53) Sun, S.; Wu, P. From Globules to Crystals: A Spectral Study of Poly (2-Isopropyl-2-Oxazoline) Crystallization in Hot Water. *Phys. Chem. Chem. Phys.* **2015**, *17* (48), 32232–32240.

(54) Rettler, E. F.-J.; Lambermont-Thijs, H. M. L.; Kranenburg, J. M.; Hoogenboom, R.; Unger, M. V; Siesler, H. W.; Schubert, U. S. Water Uptake of Poly (2-N-Alkyl-2-Oxazoline) s: Influence of Crystallinity and Hydrogen-Bonding on the Mechanical Properties. *J. Mater. Chem.* **2011**, *21* (43), 17331–17337.

(55) Li, T.; Tang, H.; Wu, P. Remarkable Distinctions in the Heat-Induced Phase Transition Processes of Two Poly (2-Isopropyl-2-Oxazoline)-Based Mixed Aqueous Solutions. *Soft Matter* **2015**, *11* (15), 3046–3055.

(56) Moffitt, M. G. Self-Assembly of Polymer Brush-Functionalized Inorganic Nanoparticles: From Hairy Balls to Smart Molecular Mimics. *J. Phys. Chem. Lett.* **2013**, *4* (21), 3654–3666.

(57) Fernandes, N. J.; Koerner, H.; Giannelis, E. P.; Vaia, R. A. Hairy Nanoparticle Assemblies as One-Component Functional Polymer Nanocomposites: Opportunities and Challenges. *MRS Commun.* **2013**, *3* (1), 13–29. https://doi.org/10.1557/mrc.2013.9.

(58) Meyer, M.; Antonietti, M.; Schlaad, H. Unexpected Thermal Characteristics of Aqueous Solutions of Poly (2-Isopropyl-2-Oxazoline). *Soft Matter* **2007**, *3* (4), 430–431.

(2) Yi, C.; Yang, Y.; Liu, B.; He, J.; Nie, Z. Polymer-Guided Assembly of Inorganic Nanoparticles. *Chem. Soc. Rev.* **2020**, *49* (2), 465–508.

(3) Liu, K.; Nie, Z.; Zhao, N.; Li, W.; Rubinstein, M.; Kumacheva, E. Step-Growth Polymerization of Inorganic Nanoparticles. *Science (80-. ).* **2010**, *329* (5988), 197–200.

(4) He, J.; Liu, Y.; Babu, T.; Wei, Z.; Nie, Z. Self-Assembly of Inorganic Nanoparticle Vesicles and Tubules Driven by Tethered Linear Block Copolymers. *J. Am. Chem. Soc.* **2012**, *134* (28), 11342–11345.

(5) Zhang, Z.; Yang, P.; Xu, H.; Zheng, H. Surface Enhanced Fluorescence and Raman Scattering by Gold Nanoparticle Dimers and Trimers. *J. Appl. Phys.* **2013**, *113* (3), 33102.

(6) Lee, A.; Andrade, G. F. S.; Ahmed, A.; Souza, M. L.; Coombs, N.; Tumarkin, E.; Liu, K.; Gordon, R.; Brolo, A. G.; Kumacheva, E. Probing Dynamic Generation of Hot-Spots in Self-Assembled Chains of Gold Nanorods by Surface-Enhanced Raman Scattering. *J. Am. Chem. Soc.* **2011**, *133* (19), 7563–7570.

(7) Bian, Z.; Tachikawa, T.; Zhang, P.; Fujitsuka, M.; Majima, T. Au/TiO2 Superstructure-Based Plasmonic Photocatalysts Exhibiting Efficient Charge Separation and Unprecedented Activity. *J. Am. Chem. Soc.* **2014**, *136* (1), 458–465.

(8) Tao, H.; Galati, E.; Kumacheva, E. Temperature-Responsive Self-Assembly of Nanoparticles Grafted with UCST Polymer Ligands. *Macromolecules* **2018**, *51* (15), 6021–6027.

(9) Park, W. M.; Champion, J. A. Colloidal Assembly of Hierarchically Structured Porous Supraparticles from Flower-Shaped Protein–Inorganic Hybrid Nanoparticles. *ACS Nano* **2016**, *10* (9), 8271–8280.

(10) Rey, M.; Law, A. D.; Buzza, D. M. A.; Vogel, N. Anisotropic Self-Assembly from Isotropic Colloidal Building Blocks. *J. Am. Chem. Soc.* **2017**, *139* (48), 17464–17473.

(11) Min, Y.; Akbulut, M.; Kristiansen, K.; Golan, Y.; Israelachvili, J. The Role of Interparticle and External Forces in Nanoparticle Assembly. In *Nanoscience and Technology: A Collection of Reviews from Nature Journals*; World Scientific, 2010; pp 38–49.

(12) Singh, G.; Chan, H.; Baskin, A.; Gelman, E.; Repnin, N.; Král, P.; Klajn, R. Self-Assembly of Magnetite Nanocubes into Helical Superstructures. *Science (80-. ).* **2014**, *345* (6201), 1149–1153.

(13) Correa-Duarte, M. A.; Pérez-Juste, J.; Sánchez-Iglesias, A.; Giersig, M.; Liz-Marzán, L. M. Aligning Au Nanorods by Using Carbon Nanotubes as Templates. *Angew. Chemie* **2005**, *117* (28), 4449–4452.

(14) Park, J. Il; Nguyen, T. D.; de Queirós Silveira, G.; Bahng, J. H.; Srivastava, S.; Zhao, G.; Sun, K.; Zhang, P.; Glotzer, S. C.; Kotov, N. A. Terminal Supraparticle Assemblies from Similarly Charged Protein Molecules and Nanoparticles. *Nat. Commun.* **2014**, *5* (1), 1–9.

(15) Gröschel, A. H.; Löbling, T. I.; Petrov, P. D.; Müllner, M.; Kuttner, C.; Wieberger, F.; Müller, A. H. E. Janus Micelles as Effective Supracolloidal Dispersants for Carbon Nanotubes. *Angew. Chemie Int. Ed.* **2013**, *52* (13), 3602–3606.

(16) Vogel, N.; Retsch, M.; Fustin, C.-A.; del Campo, A.; Jonas, U. Advances in Colloidal Assembly: The Design of Structure and Hierarchy in Two and Three Dimensions. *Chem. Rev.* **2015**, *115* (13), 6265–6311.

(17) Song, J.; Cheng, L.; Liu, A.; Yin, J.; Kuang, M.; Duan, H. Plasmonic Vesicles of Amphiphilic Gold Nanocrystals: Self-Assembly and External-Stimuli-Triggered Destruction. *J. Am. Chem. Soc.* **2011**, *133* (28), 10760–10763.

(18) Yi, C.; Zhang, S.; Webb, K. T.; Nie, Z. Anisotropic Self-Assembly of Hairy Inorganic Nanoparticles. *Acc. Chem. Res.* **2017**, *50* (1), 12–21. https://doi.org/10.1021/acs.accounts.6b00343.

(19) Akcora, P.; Liu, H.; Kumar, S. K.; Moll, J.; Li, Y.; Benicewicz, B. C.; Schadler, L. S.; Acehan, D.; Panagiotopoulos, A. Z.; Pryamitsyn, V. Anisotropic Self-Assembly of Spherical Polymer-Grafted Nanoparticles. *Nat. Mater.* **2009**, *8* (4), 354–359.

(20) Chancellor, A. J.; Seymour, B. T.; Zhao, B. Characterizing Polymer-Grafted Nanoparticles: From Basic Defining Parameters to Behavior in Solvents and Self-Assembled Structures. ACS Publications 2019.

(21) Guo, Y.; Harirchian-Saei, S.; Izumi, C. M. S.; Moffitt, M. G. Block Copolymer Mimetic Self-Assembly of Inorganic Nanoparticles. *ACS Nano* **2011**, *5* (4), 3309–3318.

(22) Wang, B.; Li, B.; Dong, B.; Zhao, B.; Li, C. Y. Homo-and Hetero-Particle Clusters Formed by Janus Nanoparticles with Bicompartment Polymer Brushes. *Macromolecules* **2010**, *43* (22), 9234–9238.

(23) Vaia, R. A.; Maguire, J. F. Polymer Nanocomposites with Prescribed Morphology: Going beyond Nanoparticle-Filled Polymers. *Chem. Mater.* **2007**, *19* (11), 2736–2751.

(24) Glotzer, S. C.; Solomon, M. J. Anisotropy of Building Blocks and Their Assembly into Complex Structures. *Nat. Mater.* **2007**, *6* (8), 557.

(25) Li, C. Y. Anisotropy Unnecessary. *Nat. Mater.* **2009**, *8* (4), 249–250.

(26) Zhou, S.; Sakamoto, T.; Wang, J.; Sugawara-Narutaki, A.; Shimojima, A.; Okubo, T. One-Dimensional Assembly of Silica Nanospheres: Effects of Nonionic Block Copolymers. *Langmuir* **2012**, *28* (37), 13181–13188.

(27) Martin, T. B.; Seifpour, A.; Jayaraman, A. Assembly of Copolymer Functionalized Nanoparticles: A Monte Carlo Simulation Study. *Soft Matter* **2011**, *7* (13), 5952–5964.

(28) Wang, J.; Xia, H.; Zhang, Y.; Lu, H.; Kamat, R.; Dobrynin, A. V; Cheng, J.; Lin, Y. Nucleation-Controlled Polymerization of Nanoparticles into Supramolecular Structures. *J. Am. Chem. Soc.* **2013**, *135* (31), 11417–11420.

(29) He, W.-N.; Xu, J.-T. Crystallization Assisted Self-Assembly of Semicrystalline Block Copolymers. *Prog. Polym. Sci.* **2012**, *37* (10), 1350–1400.

(30) Mai, Y.; Eisenberg, A. Self-Assembly of Block Copolymers. *Chem. Soc. Rev.* **2012**, *41* (18), 5969–5985.

(31) Schmelz, J.; Schacher, F. H.; Schmalz, H. Cylindrical Crystalline-Core Micelles: Pushing the Limits of Solution Self-Assembly. *Soft Matter* **2013**, *9* (7), 2101–2107.

(32) Gilroy, J. B.; Gädt, T.; Whittell, G. R.; Chabanne, L.; Mitchels, J. M.; Richardson, R. M.; Winnik, M. A.; Manners, I. Monodisperse Cylindrical Micelles by Crystallization-Driven Living Self-Assembly. *Nat. Chem.* **2010**, *2* (7), 566–570.

(33) Wang, X.; Guerin, G.; Wang, H.; Wang, Y.; Manners, I.; Winnik, M. A. Cylindrical Block Copolymer Micelles and Co-Micelles of Controlled Length and Architecture. *Science (80-. ).* **2007**, *317* (5838), 644–647.

(34) Demirel, A. L.; Meyer, M.; Schlaad, H. Formation of Polyamide Nanofibers by Directional Crystallization in Aqueous Solution. *Angew. Chemie - Int. Ed.* **2007**, *46* (45), 8622–8624. https://doi.org/10.1002/anie.200703486.

(35) Diehl, C.; Černoch, P.; Zenke, I.; Runge, H.; Pitschke, R.; Hartmann, J.; Tiersch, B.; Schlaad, H. Mechanistic Study of the Phase Separation/Crystallization Process of Poly (2-Isopropyl-2-Oxazoline) in Hot Water. *Soft Matter* **2010**, *6* (16),

(36) Giannini, C.; Ladisa, M.; Altamura, D.; Siliqi, D.; Sibillano, T.; De Caro, L.; Diehl, C.; Černoch, P.; Zenke, I.; Runge, H.; et al. Crystallisation-Driven Self-Assembly of Poly(2-Isopropyl-2-Oxazoline)-Block-Poly(2-Methyl-2-Oxazoline) above the LCST. *Soft Matter* **2015**, *6* (19), 87. https://doi.org/10.1021/acs.macromol.7b01639.

(37) Nabiyan, A.; Biehl, P.; Schacher, F. H. Crystallization vs Metal Chelation: Solution Self-Assembly of Dual Responsive Block Copolymers. *Macromolecules* **2020**, *53* (13), 5056–5067. https://doi.org/10.1021/acs.macromol.0c00792.

(38) Rudolph, T.; Lühe, M. von der; Hartlieb, M.; Norsic, S.; Schubert, U. S.; Boisson, C.; D'Agosto, F.; Schacher, F. H. Toward Anisotropic Hybrid Materials: Directional Crystallization of Amphiphilic Polyoxazoline-Based Triblock Terpolymers. *ACS Nano* **2015**, *9* (10), 10085–10098.

(39) Eckardt, O.; Pietsch, C.; Zumann, O.; von der Lühe, M.; Brauer, D. S.; Schacher, F. H. Well-Defined $SiO_2$@ P (EtOx-stat-EI) Core–Shell Hybrid Nanoparticles via Sol-Gel Processes. *Macromol. Rapid Commun.* **2016**, *37* (4), 337–342.

(40) Stöber, W.; Fink, A.; Bohn, E. Controlled Growth of Monodisperse Silica Spheres in the Micron Size Range. *J. Colloid Interface Sci.* **1968**, *26* (1), 62–69.

(41) Wu, Z.; Beltran-Villegas, D. J.; Jayaraman, A. Development of a New Coarse-Grained Model to Simulate Assembly of Cellulose Chains Due to Hydrogen Bonding. *J. Chem. Theory Comput.* **2020**, *16* (7), 4599–4614.

(42) Starr, F. W.; Sciortino, F. Model for Assembly and Gelation of Four-Armed DNA Dendrimers. *J. Phys. Condens. Matter* **2006**, *18* (26), L347.

(43) Jorgensen, W. L.; Maxwell, D. S.; Tirado-Rives, J. Development and Testing of the OPLS All-Atom Force Field on Conformational Energetics and Properties of Organic Liquids. *J. Am. Chem. Soc.* **1996**, *118* (45), 11225–11236.

(44) Kaminski, G. A.; Friesner, R. A.; Tirado-Rives, J.; Jorgensen, W. L. Evaluation and Reparametrization of the OPLS-AA Force Field for Proteins via Comparison with Accurate Quantum Chemical Calculations on Peptides. *J. Phys. Chem. B* **2001**, *105* (28), 6474–6487.

(45) Berendsen, H. J. C.; van der Spoel, D.; van Drunen, R. GROMACS: A Message-Passing Parallel Molecular Dynamics Implementation. *Comput. Phys. Commun.* **1995**, *91* (1–3), 43–56.

(46) Berendsen, H. J. C.; Grigera, J. R.; Straatsma, T. P. The Missing Term in Effective Pair Potentials. *J. Phys. Chem.* **1987**, *91* (24), 6269–6271.

(47) Anderson, J. A.; Glaser, J.; Glotzer, S. C. HOOMD-Blue: A Python Package for High-Performance Molecular Dynamics and Hard Particle Monte Carlo Simulations. *Comput. Mater. Sci.* **2020**, *173*, 109363.

(48) Park, J.-S.; Akiyama, Y.; Winnik, F. M.; Kataoka, K. Versatile Synthesis of End-Functionalized Thermosensitive Poly (2-Isopropyl-2-Oxazolines). *Macromolecules* **2004**, *37* (18), 6786–6792.

(49) Katsumoto, Y.; Tsuchiizu, A.; Qiu, X.; Winnik, F. M. Dissecting the Mechanism of the Heat-Induced Phase Separation and Crystallization of Poly (2-Isopropyl-2-Oxazoline) in Water through Vibrational Spectroscopy and Molecular Orbital Calculations. *Macromolecules* **2012**, *45* (8), 3531–3541.

(50) Demirel, A. L.; Meyer, M.; Schlaad, H. Formation of Polyamide Nanofibers by Directional Crystallization in Aqueous Solution. *Angew. Chemie Int. Ed.* **2007**, *46* (45), 8622–8624.

(51) Meyer, M.; Schlaad, H. Poly (2-Isopropyl-2-Oxazoline)– Poly (l-Glutamate) Block Copolymers through Ammonium-Mediated NCA Polymerization. *Macromolecules* **2006**, *39* (11), 3967–3970.

(52) Li, T.; Tang, H.; Wu, P. Molecular Evolution of Poly(2-Isopropyl-2-Oxazoline) Aqueous Solution during the Liquid-Liquid Phase Separation and Phase Transition Process. *Langmuir* **2015**, *31* (24), 6870–6878. https://doi.org/10.1021/acs.langmuir.5b01009.

(53) Sun, S.; Wu, P. From Globules to Crystals: A Spectral Study of Poly (2-Isopropyl-2-Oxazoline) Crystallization in Hot Water. *Phys. Chem. Chem. Phys.* **2015**, *17* (48), 32232–32240.

(54) Rettler, E. F.-J.; Lambermont-Thijs, H. M. L.; Kranenburg, J. M.; Hoogenboom, R.; Unger, M. V; Siesler, H. W.; Schubert, U. S. Water Uptake of Poly (2-N-Alkyl-2-Oxazoline) s: Influence of Crystallinity and Hydrogen-Bonding on the Mechanical Properties. *J. Mater. Chem.* **2011**, *21* (43), 17331–17337.

(55) Li, T.; Tang, H.; Wu, P. Remarkable Distinctions in the Heat-Induced Phase Transition Processes of Two Poly (2-Isopropyl-2-Oxazoline)-Based Mixed Aqueous Solutions. *Soft Matter* **2015**, *11* (15), 3046–3055.

(56) Moffitt, M. G. Self-Assembly of Polymer Brush-Functionalized Inorganic Nanoparticles: From Hairy Balls to Smart Molecular Mimics. *J. Phys. Chem. Lett.* **2013**, *4* (21), 3654–3666.

(57) Fernandes, N. J.; Koerner, H.; Giannelis, E. P.; Vaia, R. A. Hairy Nanoparticle Assemblies as One-Component Functional Polymer Nanocomposites: Opportunities and Challenges. *MRS Commun.* **2013**, *3* (1), 13–29. https://doi.org/10.1557/mrc.2013.9.

(58) Demirel, A. L.; Meyer, M.; Schlaad, H. Formation of Polyamide Nanofibers by Directional Crystallization in Aqueous Solution. *Angew. Chemie Int. Ed.* **2007**, *46* (45), 8622–8624.

(59) Meyer, M.; Antonietti, M.; Schlaad, H. Unexpected Thermal Characteristics of Aqueous Solutions of Poly (2-Isopropyl-2-Oxazoline). *Soft Matter* **2007**, *3* (4), 430–431.